\documentclass[aps,prl,superscriptaddress,amsfonts,amsmath,amssymb,reprint,showpacs,floatfix]{revtex4-1}

\usepackage{url}
\usepackage{bm}
\usepackage{graphicx}
\usepackage{amsmath}
\usepackage{amstext}
\usepackage{amssymb}
\usepackage{amsfonts}
\usepackage{amsbsy}
\usepackage{verbatim}
\usepackage{color}
\usepackage[colorlinks=true, urlcolor=blue, linkcolor=blue, citecolor=blue, pdftex]{hyperref}
\usepackage{multirow}
\usepackage{pdfpages}
\usepackage{floatrow}
\usepackage{textcomp}

\newcommand{\bs}[1]{\boldsymbol{#1}}

\begin{document}

\title{Evolution of superconducting gap anisotropy in hole-doped 122 iron pnictides}

\author{Christian Platt}
\email[]{chplatt@stanford.edu}
\affiliation{Department of Physics, Stanford University, Stanford, CA 94305, USA}
\author{Gang Li}
\email[]{li@ifp.tuwien.ac.at}
\affiliation{Institute of Solid State Physics, Vienna University of Technology, A-1040 Vienna, Austria}
\author{Mario Fink}
\email[]{mario.fink@physik.uni-wuerzburg.de}
\affiliation{Institute for Theoretical Physics and Astrophysics,
  Julius-Maximilians University of W\"urzburg, Am Hubland, D-97074
  W\"urzburg, Germany}
\author{Werner Hanke}
\email[]{hanke@physik.uni-wuerzburg.de}
\affiliation{Institute for Theoretical Physics and Astrophysics,
  Julius-Maximilians University of W\"urzburg, Am Hubland, D-97074
  W\"urzburg, Germany}
\author{Ronny Thomale}
\email[]{rthomale@physik.uni-wuerzburg.de}
\affiliation{Institute for Theoretical Physics and Astrophysics, Julius-Maximilians University of W\"urzburg, Am Hubland, D-97074 W\"urzburg, Germany}
\date{\today}

\begin{abstract}
  Motivated by recent experimental findings, we investigate the evolution 
  of the superconducting gap anisotropy in 122 iron pnictides as a function of hole doping. Employing
  both a functional and a weak coupling renormalization group approach (FRG and WRG), 
  we analyse the Fermi surface instabilities of an
  effective 122 model band structure at different hole dopings $x$, and
  derive the gap anisotropy from the leading superconducting
  instability. In the transition regime from collinear magnetism to
  $s_\pm$-wave, where strong correlations are present, we employ FRG to
  identify a non-monotonous change of the gap anisotropy in qualitative
  agreement with new experimental findings. From the WRG, which is
  asymptotically exact in the weak coupling limit, we find an
  $s_\pm$-wave to $d$-wave transition as a function of hole doping,
  complementing previous findings from FRG [Thomale et al.,
  Phys. Rev. Lett. {\bf 107}, 117001 (2011)]. The gap anisotropy of
  the $s_\pm$-wave monotonously increases towards the transition to $d$-wave
  as a function of $x$.
\end{abstract}


\maketitle

{\it Introduction.} The iron pnictides have established a new arena of
high-temperature superconductors with a remarkable variety of
structural and chemical material compositions~\cite{hoso,igor,RevModPhys.83.1589,Wang200,chubu}. Among them, the
BaFe$_{2}$As$_{2}$ (Ba-122) parent compound has received particular attention due
to its crystal quality and amenability to chemical substitution. A
$T_c$ up to 38 K at optimal doping $x\sim 0.4$ has been accomplished in Ba-122~\cite{PhysRevLett.101.107006} by
replacing Ba$^{2+}$ with K$^{+}$ of similar atomic radius to give
K$_{x}$Ba$_{1-x}$Fe$_{2}$As$_{2}$ (KBa-122). As for moderately doped KBa-122 and most other
pnictide families, experimental evidence combined with theoretical
modelling tends to be consistent with an extended $s$-wave
($s_\pm$-wave) superconducting state~\cite{PhysRevLett.101.057003}
driven by electronic correlations. Here, the superconducting gap function takes opposite
signs on hole pockets located at $\Gamma$ and $M$ versus electron
pockets at $X$ and $X'$ in the unfolded Brillouin zone with one Fe
atom per unit cell. (The existence of an $M$ hole Fermi pocket is one
of the few both significant and non-universal features of iron
pnictide materials. If present as for KBa-122, it is vital to
understanding the fundamental character of the superconducting state~\cite{PhysRevB.86.180502}.) While the $s_\pm$ state still resides in the
$A_{1g}$ lattice representation and as such cannot be distinguished
from a trivial $s$-wave in this respect, the sign change between
electron and hole pockets allows to take advantage of collinear
$(\pi,0)/(0,\pi)$ spin fluctuations as a central driver for
superconductivity~\cite{RevModPhys.84.1383, PhysRevB.78.134512,PhysRevLett.101.087004}. 

In the strong hole doping limit of KBa-122, there is conflicting
experimental evidence suggesting a $d$-wave superconducting order
parameter. 
While penetration depth~\cite{PhysRevB.82.014526} and nuclear
quadrupole resonance~\cite{doi:10.1143/JPSJ.78.083712} measurements 
only hint at a nodal superconducting state which could also imply a
nodal $s_\pm$ state, evidence in favor of $d$-wave has been deduced from thermal
conductivity~\cite{PhysRevLett.104.087005,PhysRevLett.109.087001}. The
enhanced experimental interest was preceded by the theoretical proposal of an extended $d$-wave ($d_\pm$-wave)
state~\cite{PhysRevB.80.180505} for strongly hole doped pnictides and a
concise material prediction of a $d_\pm$-wave state for
KBa-122~\cite{PhysRevLett.107.117001}. Specific heat measurements 
in K-122~\cite{louispressure,PhysRevB.87.180507} and heat transport in
RbFe$_2$As$_2$~\cite{PhysRevB.91.024502} show signatures
of a nodal gap that, upon pressure, undergoes a phase transition into
a nodeless gap~\cite{PhysRevB.91.054511}. Contrasting the thermal
conductivity profile against K-122, optimally doped KBa-122 as a
candidate for hosting a nodeless $s_\pm$-wave state and the 1111
pnictide LaFePO as a candidate for an accidentally nodal $s_\pm$
state~\cite{PhysRevB.79.224511,PhysRevLett.106.187003} show strikingly
different transport behaviour~\cite{0953-2048-25-8-084013}. This supports
the unique, possibly $d$-wave character of superconductivity in K-122,
while a recent thermal transport study is challenging previous
interpretations in favor of protected gap nodes in
K-122~\cite{0256-307X-32-12-127403}. Additionally, recent penetration
depth experiments are interpreted in favor of an $s_\pm$-wave state
for arbitrary hole doping~\cite{ruslan}, while a $d$-wave state could
in principle hardly be distinguished from a nodal $s_\pm$-wave state for
large hole doping. Moreover, heat capacity and thermal expansion experiments
appear to suggest the absence of nodes in the superconducting state
even for large hole doping~\cite{meingast}.
As another possible objection against $d$-wave in strongly hole doped KBa-122,
findings from laser ARPES seem to rule out nodes on the hole
pockets~\cite{Shimojima564,PhysRevB.89.081103}. (A particular type of
nodal $s$-wave solution preferred by interactions of small momentum scattering would be consistent with
this observation~\cite{PhysRevB.85.014511}, while the fragility of the
SC phase against disorder speaks against an $s$-wave
state~\cite{PhysRevLett.109.087001}.) Furthermore, a vortex
lattice analysis questions the existence of any vertical line nodes,
which would be a shared feature of both nodal $s_\pm$-wave and
$d$-wave order~\cite{PhysRevB.84.024507}.

Together with the question of $s_\pm$-wave vs. $d$-wave for strong
hole doping, recent heat transport revived the interest in the momentum dependence of
the superconducting gap in the underdoped regime~\cite{louisnew}. The
ratio of residual thermal resistivity in the
superconducting and the normal state allows
to resolve the change of density in the low-energy thermal transport
regime which, by assuming an only slowly changing gap amplitude,
correlates with the degree of gap anisotropy.
From an experimental perspective, ARPES sometimes allows to
resolve some $k$-dependence of the superconducting gap along the Fermi
surface, while the decreased accuracy for large $k$ as well as disorder is mostly
preventing a detailed resolution of e.g. the electron pocket gaps in
the pnictides. In principle, STM is an ideal method to resolve the
momentum dependence of the gap. For pnictides such as LiFeAs which
cleave at an electrically neutral surface, this provided a detailed
resolution of the gap function~\cite{Allan563}, showing LiFeAs to be
of moderately anisotropic $s$-wave type. (Astonishingly, in part due to the
absence of magnetic order, LiFeAs was one the most difficult pnictides
to be analysed in theory, and only at a comparatively late stage was
found to host an $s_\pm$ state~\cite{PhysRevB.84.235121}.) 
Since 122 pnictides do not cleave at a neutral plane and early ARPES data
did not observe detailed gap modulation in KBa-122~\cite{0295-5075-85-6-67002}, one had
to resort to integrated
measures of the gap function such as by thermal
Hall~\cite{PhysRevB.86.180502} or thermal
conductivity~\cite{PhysRevB.80.140503}, from the beginning.

\begin{figure*}
\includegraphics[width=1.0\columnwidth]{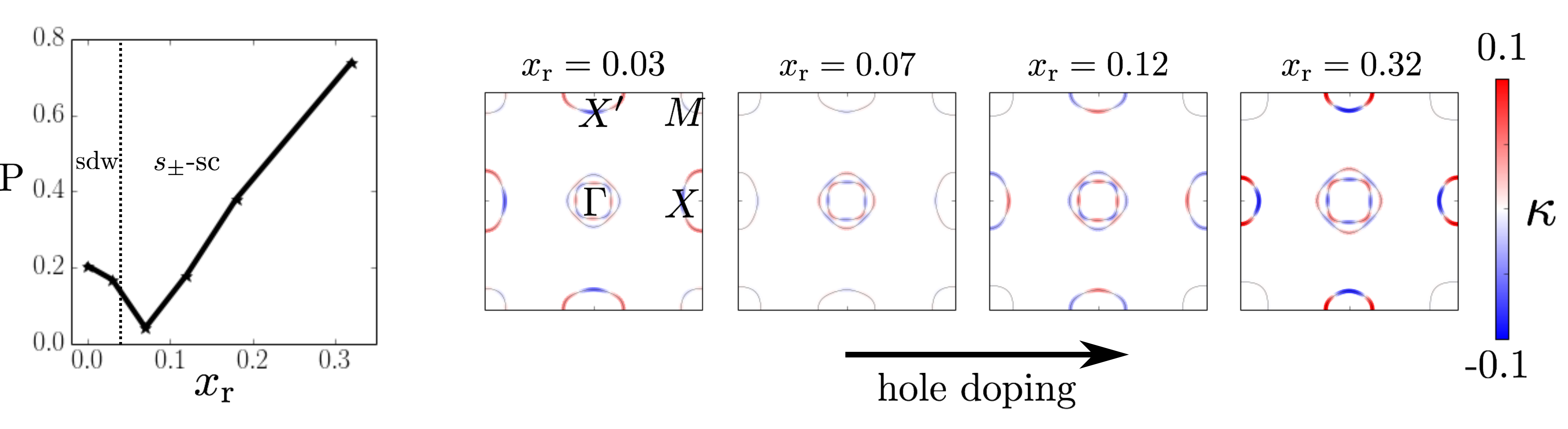}
\caption{\label{fig:fig1} (Color online) Evolution of $s_{\pm}$-wave gap
anisotropy computed from FRG in the vicinity of the magnetic phase. $x_{\text{r}}$
parametrizes the relative hole doping with respect to the center of
the $(\pi,0)/(0,\pi)$ collinear magnetic domain. Starting from
$x_{\text{r}}<0.05$ where $s_\pm$-wave is still a subleading
instability, the gap anisotropy $P$ which by Eq.~\ref{p} is the
integrated square of the gap anisotropy $\kappa$ defined in Eq.~\ref{k} first decreases as
a function of $x_{\text{r}}$, and then increases again deeper in the
superconducting phase. This feature can be
reconciled by a changing orbital-sensitive balance of competing scattering channels $\Gamma \leftrightarrow X$
and $M \leftrightarrow X$ vs. 
$X\leftrightarrow X'$ as a function of $x_{\text{r}}$.}
\end{figure*}

\begin{figure*}
\includegraphics[width=1.0\columnwidth]{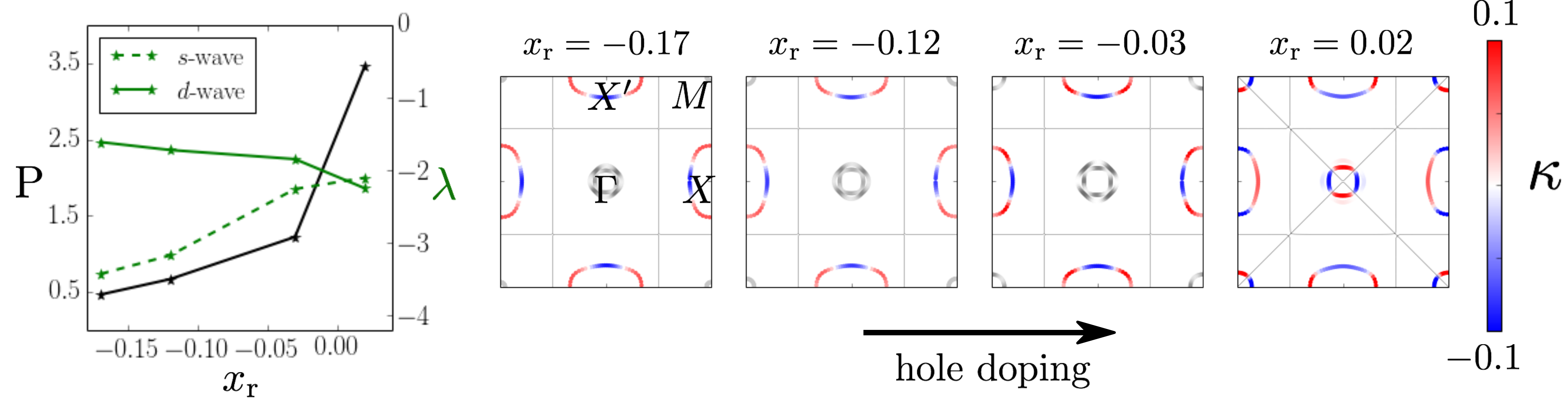}
\caption{\label{fig:fig2} (Color online) Evolution of the
  superconducting gap anisotropy computed from WRG in the vicinity of
  the $s_\pm$-wave to $d$-wave phase transition. $x_{\text{r}}$
  parametrizes the relative hole doping with respect to the transition
point. The attractive SC eigenvalue $\lambda_{d}$ (green straight line) eventually moves below
$\lambda_{s_\pm}$ (green dashed line) as a function of hole
doping. Approaching the transition from the $s_\pm$ side
($x_{\text{r}}<0$), the Fermi surface-resolved gap anisotropy $\kappa$ predominantly changes on the electron pockets (coloured) while the gap
anisotropy on the hole pockets (grey) stays unchanged. The total $s_\pm$-wave gap
anisotropy $P$ increases monotonously, and particularly steeply close to
the $d$-wave transition. We find an extended $d$-wave state with no
sign change between $\Gamma$ and $M$ hole pockets, in line with
previous calculations from
FRG~\cite{PhysRevB.80.180505,PhysRevLett.107.117001}. Beyond the
transition, the extended $d$-wave state trivially maximizes the gap anisotropy for a given gap amplitude.}
\end{figure*}

In this article, we analyse the evolution of superconducting gap anisotropy
in hole-doped KBa-122. Employing both a functional and a weak
coupling renormalization group study (FRG~\cite{RevModPhys.84.299,doi:10.1080/00018732.2013.862020} and WRG~\cite{PhysRevLett.15.524,PhysRevB.81.224505}), we compute the momentum dependence of
the gap function, starting from the coexistence phase with
collinear magnetism around half filling up to K-122 at maximum hole
doping. Around half filling, where the degree of correlations is relatively high,
FRG appears as a better choice than e.g. the random phase approximation (RPA) to
resolve the interdependencies of the particle-hole and
particle-particle parquet channels. In line with recent experiments~\cite{louisnew},
we find a non-monotonous evolution of the gap anisotropy which, as a
function of hole doping, reaches
a minimum in the superconducting phase from which on the gap keeps
increasing (Fig.~\ref{fig:fig1}). In the strong hole doping regime, we
employ WRG which provides analytically exact results in the limit of
infinitesimal interactions. In qualitative agreement with previous FRG
studies~\cite{PhysRevLett.107.117001}, we find a transition from
$s_\pm$-wave to $d_\pm$-wave~\cite{PhysRevB.80.180505}, which allows
us to study the $s_\pm$ gap anisotropy evolution as we approach the
transition point (Fig.~\ref{fig:fig2}).


{\it Superconducting gap function.}
In order to investigate the competing orders of Ba$_{1-x}$K$_x$Fe$_2$As$_2$, 
we start out from a five iron $d$-orbital description obtained by
Graser \textit{et al}.~\cite{PhysRevB.81.214503} for the undoped
parent compound. (Note that the FRG was previously extended to
additionally include the As $p$-bands~\cite{PhysRevB.89.214514}, and
found an, in principle, similar result
to the effective 5-band description.)
The corresponding 
tight-binding model is given by
\begin{equation}
H_0 = \sum_{\mathbf{k},s}\sum_{a,b}c^{\dagger}_{\mathbf{k}as}K_{ab}(\mathbf{k})c^{\phantom{\dagger}}_{\mathbf{k}bs},
\end{equation}
with $\mathbf{k}$, $(a,b)$, $s$ denoting momentum, orbital, and spin
degrees of freedom and $K_{ab}$ standing for the orbital matrix element . Except for certain
minor details, the full \textit{ab-initio} band structure for
BaFe$_2$As$_2$ at low energies is accurately reproduced by this five
orbital description~\cite{PhysRevB.81.214503}.  The different doping
levels, however, are only modeled by a rigid band shift plus mass renormalization, which for $x=0.5$
potassium replacement amounts to $0.1eV$, roughly $2.5\%$ of the
bandwidth. The filling is thereby given by $n=6.0 - x/2$ electrons per
iron atom. In order to provide a quantitatively accurate doping
evolution in such a band structure, it would be necessary to go beyond
such a rigid band approximation. Instead, we take on a qualitative
view in the following, and concentrate on the two important transition regimes as a function of
hole doping, i.e. from collinear magnetism to $s_\pm$-wave for the
underdoped case and from $s_\pm$-wave to $d_\pm$-wave, which might be
experimentally observed for the overdoped case~\cite{hackl}. 

For the interaction part $H_{\text{int}}$, we use a complete set
of onsite intra- and inter-orbital Coulomb repulsion as well as Hund's
rule and pair-hopping terms
\begin{eqnarray}
H_{\text{int}} &=& \sum_i \left[U_{\text{intra}} \sum_a n_{ia\uparrow}n_{ia\downarrow} + U_{\text{inter}} \sum_{a< b,ss'} n_{ias}n_{ibs}\right. \nonumber\\
&&\left. - J_{\text{H}}\sum_{a< b} \vec{S}_{ia}\vec{S}_{ib} 
+ J_{\text{pair}}  \sum_{a<b}c^{\dagger}_{ia\uparrow}c^{\dagger}_{ia\downarrow}c^{\phantom{\dagger}}_{ib\downarrow}c^{\phantom{\dagger}}_{ib\uparrow}\right]
\label{hint}
\end{eqnarray}
with $U_{\text{intra}} = 4.0eV$, $U_{\text{inter}} = 2.0eV$, and
$J_{\text{H}} = J_{\text{pair}} = 0.7eV$~\cite{Miyake2010JPSJ, werner12natphys331}. 
Both the functional RG and the weak-coupling RG provide an effective low-energy theory $H^{\Lambda}$
which reveals a hierarchy of favoured Fermi surface instabilities in
all channels. (Note that only the relative and not the absolute
strengths of the interaction terms matter for WRG, in which the
absolute scale is taken to the infinitesimal limit.) Referring to the
literature for more details on the methodology (for a review on
multi-orbital FRG see~\cite{doi:10.1080/00018732.2013.862020}, for the
WRG see~\cite{PhysRevB.81.224505,PhysRevB.88.064505}), for the subsequent
discussion of the superconducting gap anisotropy, it is only relevant
to appreciate that FRG and WRG adopt an appropriate band basis
$\gamma^{\dagger}_{k}$, $\gamma^{\phantom{\dagger}}_{k}$ that
diagonalizes the quadratic part of the Hamiltonian. 
(The label $k$ comprises momentum, band, and spin degrees of freedom.) The renormalized interaction in this basis yields
\begin{equation}\label{eq:effmodel}
H_{\text{int}}^{\Lambda} = \sum_{k_1,\ldots,k_4}V^{\Lambda}(k_1,k_2,k_3,k_4)\gamma^{\dagger}_{k_1}\gamma^{\dagger}_{k_2}\gamma^{\phantom{\dagger}}_{k_3}\gamma^{\phantom{\dagger}}_{k_4}.
\end{equation}
Let us particularize to the superconducting channel. The irreducible
lattice representations along with relative angular momentum of the
superconducting condensate allow to distinguish the different
superconducting orders. We decompose the pairing channel into
eigenmodes
\begin{equation}
\label{lgap}
\sum_i\oint_{FS_i} \frac{d\hat{q}}{(2\pi)v_F(q)} V^{\Lambda}(q,-q,k,-k) g_l(q) = \lambda_l g_l(k).
\end{equation}
Here, $\sum_i\oint_{FS_i}d\hat{q}$ denotes the integration along all
Fermi-surface sheets, $v_F(q)$ the Fermi velocity, and $l$ the running
index over the different eigenmodes and their corresponding
eigenvectors $g_l(k)$. Eq.~(\ref{lgap}) is form invariant to the linearized
BCS gap equation. Negative eigenvalues $\lambda_l$ signal a
superconducting instability. In weak coupling, the $\lambda_l$ convert
into a $T_{c}^l\sim \exp (-1/(N_{\text{F}}\vert\lambda_l\vert))$, where $N_F$ denotes the density of states at the Fermi level,
i.e. the most negative $\lambda_{l^*}\equiv \lambda_1$ is the dominant superconducting instability.

It is the gap form factor $g_1(k)$ that hosts the information about the momentum
dependence of the superconducting gap. To create a local measure $\kappa_i(k)$ 
for the gap anisotropy, we consider the deviation of  $g_1(k)$ with respect to its mean on the respective Fermi-surface sheet
\begin{equation}
\kappa_i(k) = g_1(k) - \oint_{FS_i}\frac{d\hat{q}}{2\pi}g_1(q), \label{k}
\end{equation}
where $i$ denotes a running index over individual pockets. Note that $\kappa$ is
defined as the gap eigenvector $g_1(k)$ subtracted by its pocket
average, and as such takes positive and negative values even for an
$s$-wave eigenvector.  
The integrated square of $\kappa_i(k)$
\begin{equation}
P = \sum_i\oint_{FS_i}\frac{d\hat{q}}{2\pi}\ \kappa_i(q)^2, \label{p}
\end{equation}
then gives a reasonable measure for the total gap anisotropy.

{\it Gap anisotropy in the underdoped regime.} Within FRG, one finds a
second-order phase transition from a collinear $Q=(\pi,0)/(0,\pi)$ spin
density wave instability to an $s_\pm$ superconducting instability in
the underdoped regime. The SDW form factor is nodal~\cite{PhysRevB.79.014505}, mainly reflecting the
change of $d_{xz}$ and $d_{yz}$ orbital weight on electron and hole pockets.\\ 
What is experimentally perceived as the coexistence regime of magnetism and superconductivity
is reconciled in FRG by the domain where the leading eigenvalue in the
pairing channel $\lambda^{\text{SC}}$ is smaller, but in close
proximity to the leading eigenvalue in the magnetic (i.e. crossed
particle-hole) channel $\lambda^{\text{SDW}}$. As opposed to RPA
approaches, where a systematic discussion of sub-leading instabilities
in different parquet channels
is impossible due to the lack of vertex corrections between different
channels, the FRG allows us to analyze the dominant subleading
superconducting state at the onset of SDW order. \\
Fig.~\ref{fig:fig1} defines a relative doping level $x_{\text{r}}$ where $x_{\text{r}}=0$
is the center of the SDW regime. Plotting $P(x_{\text{r}})$ reveals a
reduction of gap anisotropy setting in already in the SDW-dominated
regime and continuing to the $s_\pm$-wave regime. The main change of
gap anisotropy $\kappa$ is observed for the electron pockets. Starting
from $x_{\text{r}}=0$, the background of SDW order naturally explains
an enhanced tendency for gap anisotropy and its reduction as we are
leaving the SDW regime for $x_{\text{r}}>0$. This trend matches the
experimental observation~\cite{louisnew}.\\ 
In related theoretical works, it was shown that gap nodes in the superconductor can be imposed due to
antiferromagnetism~\cite{PhysRevB.80.100508,PhysRevB.85.144527}. Intuitively, this is
found by assuming a mean field description of the background
SDW order and by considering the onset of superconductivity for the
reconstructed Fermi surface~\cite{PhysRevB.88.064505}.
Assuming
a scenario dominated by intra-orbital interactions, the gap anisotropy
evolution as
a function of hole doping can
be understood by analyzing the orbital-sensitive scattering channels
between hole and electron pockets according to $\Gamma \leftrightarrow
X/X'$ and $M \leftrightarrow X/X'$ vs. the electron-electron pocket
scattering along $X\leftrightarrow X'$. In general, the
$X\leftrightarrow X'$ scattering enhances the electron anisotropy in
order to minimize the energy penalty from repulsive interactions. By
contrast, the $\Gamma \leftrightarrow
X/X'$ scattering also induces anisotropy inherited from the change of
$d_{xz}/d_{yz}$ orbital content along the Fermi surface, but tends to
drive a homogenous gap on $X/X'$ for connected Fermi surface pieces of equal
orbital content~\cite{PhysRevLett.106.187003,hankeplus}. This competition is not
particularly modified for a small change of $x_{\text{r}}$. The crucial
effect derives from the change of $d_{xy}$ orbital content of the
electron pockets as a function of hole doping, and, as such, the relevance of 
$M \leftrightarrow X/X'$ scattering from the $M$ hole pockets which is
of solely $d_{xy}$ orbital content. As the electron pockets shrink due
to hole doping, their range of $d_{xy}$ orbital content is reduced and
concentrates on a small Fermi surface slice centered around the $\Gamma-X/X'$ front
tips of the electron pockets~\cite{PhysRevLett.106.187003}. This evolution eventually even yields an
inversion of the electron pocket anisotropy which can be observed in
Fig.~\ref{fig:fig1} for $x_{\text{r}}=0.32$. The general trend for the
overall gap anisotropy P, however, stays monotonous in this regime of $x_{\text{r}}$.

{\it Gap anisotropy at the s-wave to d-wave transition.} Guided by
recent experimental evidence~\cite{hackl}, a transition from $s$-wave
to $d$-wave in KBa-122 may occur in the strong hole doping
regime.
Note that while self-energy effects seem to be less important for
strong hole doping, mass renormalization appears to become even more
significant, and that spin
fluctuations, even though incommensurate, might persist
further in hole-doped than electron-doped 122
compounds~\cite{PhysRevLett.106.067003}. This is consistent with the
finding from correlation-induced mass
enhancements~\cite{PhysRevLett.107.166402}, a claimed improvement of
experimental evidence and theoretical modelling through a DMFT+DFT
analysis~\cite{gabi,1367-2630-16-8-083025}, and a detailed ARPES
analysis at optimal doping~\cite{PhysRevB.86.144515}. (Recently,
the enlarged density of states close but not at the Fermi level has also
been suggested to explain the enhancement of correlations in
K-122~\cite{PhysRevB.92.144513}). From NMR, strong spin and charge
fluctuations in K-122 are found close to
criticality~\cite{PhysRevB.93.085129}.)  Early ARPES data for K-122
found only hole pockets present~\cite{PhysRevLett.103.047002}, as
confirmed by de Haas van Alphen
measurements~\cite{doi:10.1143/JPSJ.79.053702}.  More recent ARPES data for
$x=0.9$~\cite{PhysRevB.88.220508} identified an according Lifshitz
transition to occur around $x=0.7-0.9$, which from density functional
theory has been located at $x\sim
0.9$~\cite{PhysRevLett.112.156401}. The dominant hole pocket at $M$ 
(in the unfolded zone) motivated previous FRG studies to predict $d$-wave
in K-122 at an early stage~\cite{PhysRevLett.107.117001}. As a general
tendency, which was also confirmed by later RPA
studies~\cite{PhysRevB.84.144514}, the propensity of forming $d$-wave
should become increasingly competitive to $s$-wave as a function of
hole doping, a notion which recently is also confirmed by Raman
spectroscopy~\cite{PhysRevX.4.041046,rudi}.

In order to adopt an approach for this question of $d$-wave in KBa-122,
which is analytically exact in the weak coupling limit, we
have expanded the original WRG to a multi-band / multi-pocket scenario
with orbital-sensitive interactions according to Eq.~\eqref{hint}. We
indeed find a phase transition from $s_\pm$-wave to $d_\pm$-wave
superconductivity. The precise doping where the transition occurs,
however, sensitively depends on the details of the interactions and
Fermiology, and occurs at lower hole doping than the experimentally
promising region. This is not surprising. First, the limit of
infinitesimal interactions oversimplifies the degree to which the spin
fluctuations are able to affect the superconducting pairing by only
considering diagrams quadratic in the Hubbard scale $U$. Second,
disorder present in the measured samples could significantly modify
the competition of $s$ vs $d$-wave where, for example, an accidentally nodal
$s_\pm$-wave state can be rendered gapped~\cite{PhysRevB.79.094512}. 

Nevertheless, we wish to investigate on analytically controlled footing how the
$s_\pm$-wave gap anisotropy evolves in proximity to the $s/d$-wave
phase transition. For the given relative interaction strengths
described below Eq.~\eqref{hint}, we employ $x_{\text{r}}$ as a
relative hole doping parameter where $x_{\text{r}}=0$ defines the
$s/d$-wave phase transition. The evolution graph of $\lambda_{s_\pm}$
vs. $\lambda_{d_{\pm}}$ is depicted in Fig.~\ref{fig:fig2}, along with
the monotonously increasing gap anisotropy $P(x_{\text{r}})$ which is
particularly steeply increasing in immediate vicinity of
$x_{\text{r}}=0$. The strongest change in $\kappa$ is observed for the
electron pockets while the hole pocket anisotropy (shaded grey) is
hardly changed. The electron pocket gap anisotropy in the gapped
$s_\pm$-wave phase (the $s_\pm$-wave nodal lines marked by straight
thin lines do not intersect the Fermi pockets) increases until the
transition occurs into the $d_\pm$-wave state. As predicted
in~\cite{PhysRevB.80.180505}, this state features $d_{x^2-y^2}$-wave type sign changes
on individual pockets along with an additional sign change from
hole to electron pockets, rendering the leading harmonic contribution
to be $\Delta_{d_\pm}(\bs{k}) \sim \cos(2k_x)-\cos(2k_y)$. \\
Similar to $s$-wave vs $s_\pm$-wave, the irreducible lattice representation for both $d_{x^2-y^2}$-wave and
$d_{\pm}$-wave is always $B_{1g}$. While $d_{x^2-y^2}$-wave
and $d_{\pm}$-wave would in principle be viable candidates for
unconventional superconducting instabilities at weak coupling, we have
never observed a leading $d_{x^2-y^2}$-wave in our investigations of
KBa-122. Note, that in comparison to the FRG result, where the driving
mechanism for $d$-wave was the intra-pocket scattering around $M$, the
$X\leftrightarrow X'$ alone seems to be strong enough for
infinitesimal coupling to make $d_{\pm}$-wave become favorable over
$s_\pm$-wave. \\
The universal enhancement of the $s_\pm$ gap anisotropy at
the transition to $d_\pm$ is a relevant observation in terms of the
possible stabilization of an $s+id$-wave superconducting state in the
iron pnictides. The $s+id$-wave state has been analyzed from
Ginzburg-Landau theory~\cite{PhysRevLett.102.217002} and was microscopically predicted to emerge in
iron pnictides~\cite{PhysRevB.85.180502, PhysRevLett.108.247003}. The central driver for such a non-chiral
superconducting state, which still breaks time-reversal symmetry and
$D_4$ lattice symmetry, is the maximisation of condensation
energy. Now, having a close-to-nodal $s_\pm$-wave gap and a
symmetry-protected nodal $d$-wave gap at the $s/d$-wave transition in
the pnictides, yields a strong condensation energy gain both by
removing the $d$-wave nodes and by reducing the $s_\pm$ gap
anisotropy. This would make $s+id$-wave particularly favorable energetically. Another $s$-wave state with unequal
hole pocket signs was proposed to induce a time-reversal symmetry
breaking $s+is$ state~\cite{PhysRevB.87.144511}. From WRG, we have not
found an indication for this formation so far.




{\it Conclusion.} We have analysed the evolution of the
superconducting gap for the hole-doped KBa-122 iron pnictides. In the
underdoped regime, our FRG analysis qualitatively reproduces the experimental
finding of an enhanced gap anisotropy in the proximity to collinear
magnetism. This anisotropy first decreases and then
monotonously increases as a function of deeper hole doping into the
$s_\pm$-wave phase. For the analytically controlled limit of
infinitesimal interactions, our weak coupling RG analysis gives an
$s_\pm$-wave to $d_\pm$-wave transition, which might relate to the
experimental regime of strong hole-doping regime around K-122. The
$s_\pm$-wave universally increases monotonously towards the
transition, and as such provides support for the possible formation of
an $s+id$-wave state in the transition regime.\\
From a broader perspective, it would be interesting to apply our
approach to electron-doped pnictides as well. 
In addition, isovalent P-doping, implying an energy shift of the hole
pocket at the $M$ point~\cite{PhysRevLett.106.187003}, might crucially
change the scenario from one studied here. Experimentally, the degree of anisotropy
for the P-doped 122 family was found to be
enhanced~\cite{Hashimoto1554,PhysRevX.2.011010}, which, in absence of
the $M$ hole pocket, is consistent
with our analysis of the $s_\pm$-wave gap anisotropy.




{\it Acknowledgments}. We thank A.~V.~Chubukov and D.~Scalapino for
discussions. We thank R.~Prozorov and L.~Taillefer for helpful
comments, and I.~I.~Mazin for a careful reading of the manuscript. 
The work was supported by the DFG (Deutsche
Forschungsgemeinschaft) through the focus program DFG-SPP 1458 on
iron-based superconductors. W. H. and R. T. acknowledge support by the
ERC (European Research Council) through
ERC-StG-TOPOLECTRICS-Thomale-336012. R.T. was supported by DFG-SFB 1170. 
CP was supported by the Leopoldina Fellowship Programme LPDS 2014-04.

\bibliographystyle{prsty}
\bibliography{refs}

\end{document}